%% file: main.tex
\documentclass[letterpaper,  journal]{IEEEtran}  % Comment this line out
\pdfoutput=1                                                        % if you need a4paper
\IEEEoverridecommandlockouts                              % This command is only

\usepackage{graphics} % for pdf, bitmapped graphics files
\usepackage{epsfig} % for postscript graphics files
\usepackage{amsmath} % assumes amsmath package installed
\usepackage{amssymb}  % assumes amsmath package installed
\usepackage{color}
\usepackage{float}
\usepackage{algorithm}
\usepackage{algorithmic}
\usepackage{calrsfs}
\usepackage{textcomp}

\DeclareMathAlphabet{\pazocal}{OMS}{zplm}{m}{n}

\begin{document}

\title{IntelligentCrowd: Mobile Crowdsensing via Multi-Agent Reinforcement Learning}

\author{Yize Chen and Hao Wang
	\thanks{Y. Chen is with the Department of Electrical and Computer Engineering, University of Washington, Seattle, WA 98195 USA,  email: yizechen@uw.edu. 
		
		H. Wang is with the Department of Data Science and Artificial Intelligence, Faculty of Information Technology, Monash University, Melbourne VIC 3800, Australia, email:  hao.wang2@monash.edu.
	
This work is in part supported by the FIT Academic Staff Funding of Monash University.

© 2020 IEEE.  Personal use of this material is permitted.  Permission from IEEE must be obtained for all other uses, in any current or future media, including reprinting/republishing this material for advertising or promotional purposes, creating new collective works, for resale or redistribution to servers or lists, or reuse of any copyrighted component of this work in other works.}
}

\maketitle

\begin{abstract}
\input{abstract}
\end{abstract}


\section{Introduction}
\input{intro}

\section{System Model and Problem Formulation}
\label{model}
\input{model}

\section{Multi-Agent Reinforcement Learning}
\label{rl_algo}
\input{rl}

\section{Performance Evaluations}
\label{results}

\input{results}

\section{Discussion and Conclusion}
\label{conclusion}
\input{conclusion}

\bibliographystyle{IEEEtran}
\bibliography{bib}
%\vspace{-20pt}
%\begin{IEEEbiography}
%	[
%	{\includegraphics[width=1in,height=1.25in,clip,keepaspectratio]{yize.jpg}}
%	]
%	{Yize Chen} received the B.S. degree in Automatic Control from Chu Kochen College, Zhejiang University, Hangzhou, China in 2016. He is currently pursuing his Ph.D.degree in Electrical Engineering at University of Washington, Seattle, WA. He held internship positions at Harvard University, Los Alamos National Laboratory, and Microsoft Research. His research is on the learning, optimization and control in power systems.
%\end{IEEEbiography}

%\begin{IEEEbiography}
%	[
%	{\includegraphics[width=1in,height=1.25in,clip,keepaspectratio]{hao.jpg}}
%	]
%{Hao Wang} (Member, IEEE) received the Ph.D. degree from The Chinese University of Hong Kong, Hong Kong, in 2016. 
%He is currently a Lecturer with the Department of Data Science and Artificial Intelligence, Faculty of Information Technology, Monash University, Melbourne, VIC, Australia. He was a Postdoctoral Research Fellow with Stanford University, Stanford, CA, USA, and a Washington Research Foundation Innovation Fellow with the University of Washington, Seattle, WA, USA. His current research interests include optimization, machine learning, data analytics, and emerging technologies (e.g., blockchain) for power and energy systems.
%\end{IEEEbiography}

\end{document}

%% file: abstract.tex
The prosperity of smart mobile devices has made mobile crowdsensing~(MCS) a promising paradigm for completing complex sensing and computation tasks. In the past, great efforts have been made on the design of incentive mechanisms and task allocation strategies from MCS platform's perspective to motivate mobile users' participation. However, in practice, MCS participants face many uncertainties coming from their sensing environment as well as other participants' strategies, and how do they interact with each other and make sensing decisions is not well understood. In this paper, we take MCS participants' perspectives to derive an online sensing policy to maximize their payoffs via MCS participation. Specifically, we model the interactions of mobile users and sensing environments as a multi-agent Markov decision process. Each participant cannot observe others' decisions, but needs to decide her effort level in sensing tasks only based on local information, e.g., her own record of sensed signals' quality. To cope with the stochastic sensing environment, we develop an intelligent crowdsensing algorithm \emph{IntelligentCrowd} by leveraging the power of multi-agent reinforcement learning~(MARL). Our algorithm leads to the optimal sensing policy for each user to maximize the expected payoff against stochastic sensing environments, and can be implemented at the individual participant's level in a distributed fashion. Numerical simulations demonstrate that \emph{IntelligentCrowd} significantly improves users' payoffs in sequential MCS tasks under various sensing dynamics.

%% file: intro.tex
Driven by the explosion of smart devices (e.g., smartphones and wearable devices), mobile crowdsensing~(MCS)~\cite{ganti2011mobile} has emerged as a promising sensing paradigm for data collection. A typical MCS system outsources small sensing tasks to a large crowd of device users to leverage sensing and computing power of mobile devices. Smart devices are often equipped with built-in sensors, including but not limited to GPS, camera, gyroscope, and accelerometer, and thus can accomplish various social and commercial tasks for different applications, such as traffic reporting, environment monitoring, social interactions, and e-business~\cite{ganti2011mobile}. In practice, mobile sensing participants face great uncertainties from the sensing environment and its interactions with the MCS service provider and other participants. In the localization task, each mobile participant senses stochastic location signals, while it is interesting to investigate how all participants can jointly complete the
sensing task. This paper aims to understand how participants can make optimal sensing decisions against uncertainties. We model the participants' interactions using Markov decision processes~(MDPs), and develop a multi-agent reinforcement learning~(MARL) algorithm-\emph{IntelligentCrowd} for all MCS participants to learn optimal sensing policies simultaneously.

%Some related work
\subsection{Related Work}
One critical issue in MCS is how to incentivize mobile users to participate in sensing programs, since performing sensing tasks will consume resources and incur costs to the participants. Therefore, great efforts have been made to design incentive mechanisms~\cite{yang2012crowdsourcing, zhang2014free,zhan2018incentive} to enroll users for MCS. For example, in \cite{yang2012crowdsourcing}, an incentive mechanism was designed using a Stackelberg game framework.
%, in which the MCS service provider is the leader and the MCS participants are the followers. A unique Stackelberg equilibrium is attained and the utility of the provider is maximized. 
In \cite{zhang2014free}, the authors considered the opportunistic nature of participants and proposed three online incentive mechanisms using a reverse auction, offering more flexibility in recruiting opportunistically encountered participants. In~\cite{zhan2018incentive}, a bargaining-based incentive mechanism was designed and a distributed iterative algorithm is used to solve the bargaining problem between the sensing platform and smart device users.  

Once mobile users are participating in MCS programs, another critical problem is how to assign tasks and then rewards to participants considering the sensing/task diversity. Existing studies have focused on task allocation from the MCS service provider ~\cite{ guo2017activecrowd, liu2016taskme, xiao2015multi,he2014toward}. For example, studies in \cite{he2014toward} showed that the optimal task allocation problem is NP-hard since sensing tasks are often associated with different locations and MCS participants under time constraints. Therefore, approximation algorithms were developed in \cite{he2014toward,li2018deep,liu2019resource} to solve a satisfactory task allocation solution with a proven approximate ratio. In \cite{guo2017activecrowd}, a worker selection framework was proposed for multi-task MCS environments with time-sensitive tasks and delay-tolerant tasks. In \cite{liu2016taskme}, the authors developed a bi-objective optimization problem to address a multi-task-oriented participant selection problem. 
%, including two situations: few participants performing more tasks and more participants working on few tasks. 
Sensing task assignment problem has also been studied in mobile social networks \cite{xiao2015multi}, and an online task assignment algorithm was designed using a greedy strategy. Recent literature also discussed the potential of applying deep reinforcement learning algorithm to the task of resource allocation and MCS~\cite{chen2018performance, huang2019deep,lei2019multiuser}, yet there has not been discussions about the interactions between different participants or rewards allocation as an environment factor.

\subsection{Motivation and Contributions}
In real-world applications, the MCS participants are facing many uncertainties that affect their decisions. For example, due to stochastic sensing environments, participants taking the same sensing strategies may lead to different sensed information. In addition, participants' economic return given by the service provider depends not only on their own efforts but also on other participants' decisions. However, most of the aforementioned studies \cite{yang2012crowdsourcing, zhang2014free,zhan2018incentive, guo2017activecrowd, liu2016taskme, xiao2015multi,he2014toward} were platform-centric, and did not consider participants' behaviors under uncertainties and fast-changing environments. Bad decision-making cannot give participants enough incentives, nor the service provider could get high-quality crowd-sourced information. Moreover, considering the interactions among users and temporal correlations of decisions, it is a challenging task to apply MCS in practice. This motivates us not to restrict ourselves in learning users' individual preferences or choices~\cite{karaliopoulos2016first}, but to study how to make sequentially optimal sensing decisions from the participants' perspectives, especially under uncertainties. 

Reinforcement learning~(RL) is a set of machine learning algorithms, which has recently been applied to solve many real-world control and decision problems~\cite{sutton1998reinforcement}. For instance, in~\cite{schulman2015trust}, a deep RL algorithm is used for robotic motion planning under stochastic environments; in~\cite{wang2017energy}, the authors proposed a reinforcement learning algorithm which maximizes the arbitrage for a single battery; in~\cite{xiao2017secure}, a single-agent RL algorithm is proposed for designing incentives for MCS participants. Yet RL has not been fully exploited under the multi-agent setting, which is appealing in many real-time problem settings such as MCS scenarios with the nature of multiple participants. 

In this paper, we model the MCS participants' behaviors using multi-agent Markov decision processes~(MDPs), in which participants make decisions on their sensing efforts under uncertainties of sensing quality and payoffs. We solve MDPs using a multi-agent reinforcement learning algorithm in an online fashion. The main contributions of this paper are summarized as follows:
\begin{itemize}
	\item \emph{User-centric MCS:} We take participants' perspective to design the optimal strategy for determining the efforts that would maximize participants' payoffs given an incentive mechanism; %The algorithm is individually rational and profitable.
	\item \emph{Online learning and decision making:} We design an online distributed MCS algorithm, namely \emph{IntelligentCrowd}, which makes use of deep reinforcement learning to learn the policy of MCS participation;
	\item \emph{Performance Evaluation:} We validate our proposed algorithm's performance under various kind of stochastic environments, which helps MCS participants get payoffs from a series of crowdsensing tasks.
\end{itemize}

The remainder of this paper is organized as follows. In Section \ref{model}, we present the MCS model. In Section \ref{rl_algo}, we formulate MDPs for mobile users and develop a multi-agent reinforcement learning algorithm. In Section \ref{results}, we show numerical simulation results. Final concluding remarks are made in Section \ref{conclusion}.

%% file: model.tex
In this section, we first model the MCS system, the value of information, and the user's effort of participation. We will then formulate the MCS problem, in which we aim to help MCS users make the participation decisions to maximize their accumulated payoffs in an uncertain sensing environment.
\begin{figure}[h]
	\centering
	\includegraphics[scale=0.25]{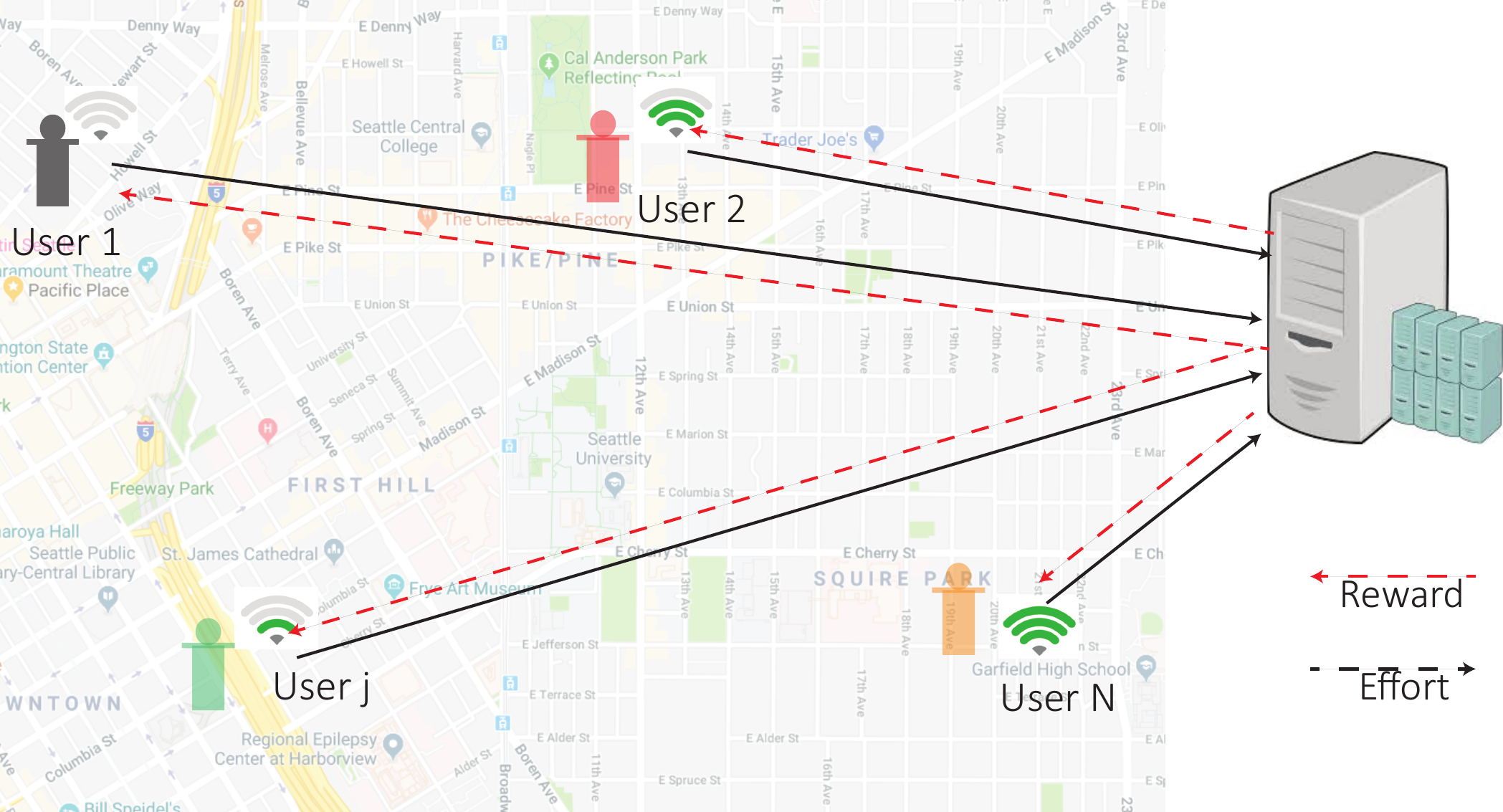}
	\caption{In an MCS system, the service provider gives rewards in exchange for information collected by MCS participants. From a participant's perspective, she is situated in a dynamic environment with time-varying sensing capabilities, and she wants to design a sensing strategy to maximize her payoffs in MCS.}
	\label{fig:intro}
\end{figure}

%\textcolor{red}{We can put the system figure here and introduce the system briefly before we dive into the modeling part.}

\subsection{MCS System}
Consider an MCS system with a set of $\mathcal{N}=\{1,...,N\}$ mobile users and a service provider. There are two-way communications between mobile users and the service provider. Each user has the capability of sensing some data in a certain area and a certain timestep~(as shown in Fig.~\ref{fig:intro}). We consider an MCS campaign over a finite and discrete-time horizon $\mathcal{T}=1,...,T$, and $N$ voluntary participants equipped with sensors perform sensing tasks for the service provider. At each time slot $t \in \mathcal{T}$, the service provider first publicizes the sensing tasks with a total reward budget $R_t$ which is time-dependent. All the participating users take their efforts to collect information and will be allocated a portion of $R_t$ based on their contribution to the overall crowdsensed information, which will be discussed later in this section.

\subsection{User Model}
We will now present the sequential user behavior in the participation of MCS along with its payoff from the service provider. 

\subsubsection{Action of Participants}
To participate in an MCS task, each user $i$ needs to select an \emph{effort level} $x_{i,t}\geq0$ and send sensed data to the service provider in time slot $t$. Performing sensing tasks will incur costs to users. At the same time, each user $i$ will also get a reward $r_{i,t}$ assigned by the service provider based on how good the quality of the sensed data is. We also adopt the notion $\textbf{x}_t=\{x_{1,t},...,x_{N,t}\}$ and $\textbf{r}_t=\{r_{1,t},...,r_{N,t}\}$ to denote the collective effort and rewards profile at each time, while we take notation ${X}_i$ as the set of possible efforts for agent $i$.

Before we present the detailed models of the costs and rewards for users, we firstly introduce the measure of the \emph{value of information}~(VoI) with respect to the user's action. We adopt the notion \emph{quality of information}~(QoI) $q_{i,t}$, which is a real-valued scalar indicating the quality of user $i$'s sensed data at time $t$. Similarly, we denote $\textbf{q}_t=\{q_{1,t},...,q_{N,t}\}$ as the set of observed QoI for all users, and $O_i$ as the set of possible QoI observations for participant $i$. Note that, due to the mobility of users and the variations of sensing tasks, $q_{i,t}$ is a stochastic process over time and often unpredictable. Then the VoI of user $i$ in time slot $t$ is based on user's contribution $q_{i,t} x_{i,t}$. We assume that after the campaign at time $t$, the past level of QoI  $\, \textbf{q}_t$ is revealed to all participants. Yet user doesn't know others' actions $\{x_{j,t}, j\neq i\}$ since everyone is making decisions independently.

\subsubsection{Payoff Function}
Participant $i$'s payoff is composed of reward from the service provider along with her sensing cost. For instance, in the task of localization service, individual participant receives reward by submitting sensing data of varying values, while incurring costs such as battery consumption and data usage. The reward $ r_{i,t}=\frac{x_{i,t} q_{i,t}}{\sum_{j\in \mathcal{N}}x_{j,t} q_{j,t}}R_t$ for user $i$ is proportional to its share of VoI at current timestep, while a participation cost is incurred based on the effort she makes $c_i x_{i,t}$ with $c_i \geq 0$. Then in sum, participant $i$'s payoff function is
\begin{equation}
\label{equ:payoff}
U_{i,t}(x_{i,t},q_{i,t})= \frac{x_{i,t} q_{i,t}}{\sum_{j\in \mathcal{N}}x_{j,t} q_{j,t}}R_t-c_i x_{i,t} .
\end{equation}

For simplicity, in this work, we assume that $R_t$ is notified by the service provider at time $t$, while $c_i$ is known to each participant. Our framework can also take the stochastic case of $R_t$ and $c_i$ into account.

\subsection{Payoff Maximization Problem}
With user's payoff designed, in this paper, we take a user-centric perspective. Our objective is to find a sequential decision $x_{i,t}$, which maximizes user $i$'s total expected discounted payoff $\sum_{t=1}^{T}\gamma^t U_{i,t}$ during the total MCS participating period, where $\gamma\in[0, \, 1]$ is a pre-defined discount factor. %When $\gamma=1$, we are simply looking the accumulated payoffs.

If the user's dynamical sensing environment $q_{i,t}$ is known or can be predicted, finding $x_{i,t}$ is a sequential decision problem. Intuitively, we can find solutions via a model-based dynamical system, which uses either the off-the-shelf offline optimization method or predictive control which maximizes \eqref{equ:payoff} with system's dynamical constraints. Based on the available information and participants' interactions, essentially for each MCS participant $i$, we can cast user's effort level as a policy mapping function from past QoI:
\begin{equation}
x_{i,t}=\pi_{\theta_i}(\mathbf{q}_{t-K},...,\mathbf{q}_t) ,
\end{equation}
where $K$ is the total window length of past QoI taken into consideration, and we adopt $\mathbf{\theta}=\{\theta_1,...,\theta_N\}$ to denote the policy function's parameters.

However, we are faced with two difficulties in applying previous-mentioned approaches to find $x_{i,t}$. Note that such payoff function is not revealed to any agent before or during training, and learning to make crowdsensing decision in an unknown environment is a challenging problem. Firstly, the participant may be situated in a highly stochastic environment. The high dimensionality of the state spaces for both users' effort levels and sensing environments, the forecast accuracy of future sensing environments both restrict the performance of existing methods. Secondly, the nature of interactions among MCS participants has made it hard to model the system dynamics accurately. For instance, one agent chooses a certain effort $x_{i,t}$, while this choice not only affects other agents' current payoff, it also could impact all agents' future decisions by considering the dynamics of sensing environments. To tackle these difficulties of modeling and decision-making, we take a machine learning approach, which automatically learns to choose effort levels for multiple participants in a stochastic QoI environment.

%% file: rl.tex
To solve the real-time payoff maximization problem for all agents, in this section, we will first describe the problem setup using MDPs, which have been developed for the discrete-time stochastic control process. Reinforcement learning algorithm, e.g., Q-Learning~\cite{sutton1998reinforcement}, could be applied for solving single-agent MDPs. Yet to solve multi-agent MDPs with non-stationary participants' sensing environment aided by partially observable information~(e.g., each participant $i$ does not know others' sensing efforts $x_{j,t},j=1,...,N, \, j\neq i$), we extend the deep reinforcement learning algorithm proposed by~\cite{lowe2017multi}, and design our MARL algorithm \emph{IntelligentCrowd}. We will then illustrate in Section~\ref{results}, under various kinds of dynamical sensing environments, \emph{IntelligentCrowd} is able to simultaneously find sensing efforts for each participant under online setting.

\subsection{Multi-Agent MDPs}

In normal MDPs models, we are only interested in the stochastic decision making by using past actions and system states. The decision on the effort level of MCS participants is a natural extension of MDPs to the multi-agent case. Mathematically, we are considering a set of actions $X_1,...,X_N$ and a set of sensed QoI level $O_1,...,O_N$. For the overall system, each step's state is simply $ O_1\times...\times O_N$. By taking a joint action for all MCS participants $X_1\times...\times X_N$, we also define two functions related to states~(QoI) and actions~(sensing effort): a) the sensing environment takes a state transition function $T:\, O_1\times...\times O_N\times X_1\times ...\times X_N \mapsto O_1\times...\times O_N$, and b) the reward~(payoff) for each agent $U_i:\, O_1\times...\times O_N \times X_i \mapsto \mathbb{R}$.

\begin{figure}[!t]
	\centering
	\includegraphics[scale=0.25]{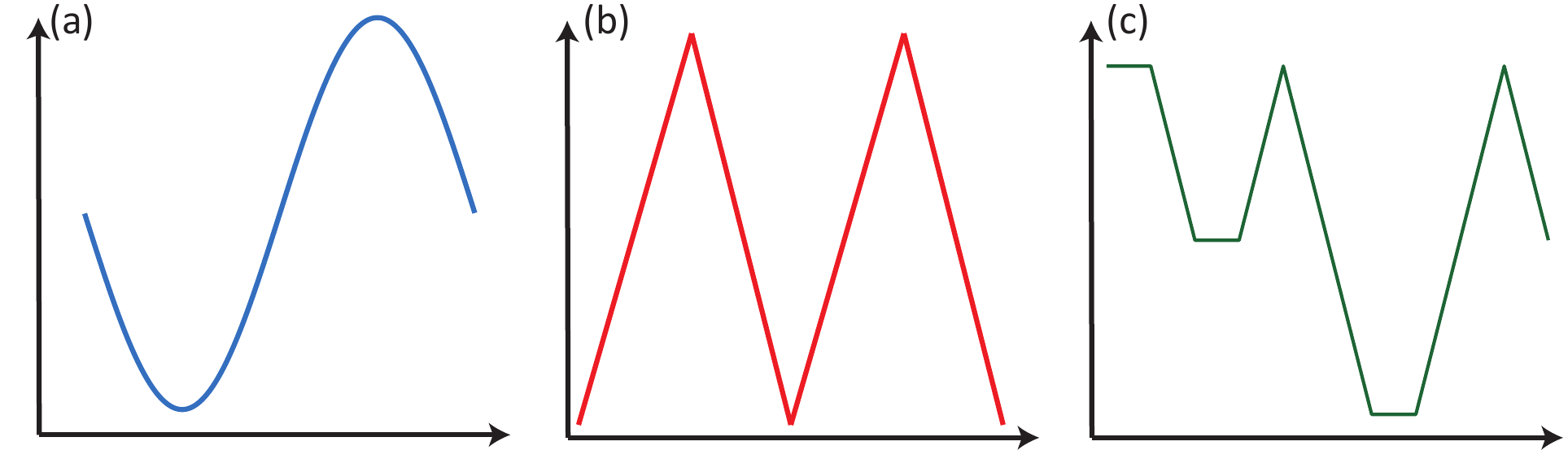}
	\caption{\small
		{
			MCS participant may be situated in various types of QoI dynamics. We consider three type of signal dynamics during simulation: (a) Sine, (b) Linear, (c) Markov chain.}
	}
	\label{fig:signals}
\end{figure}

Reinforcement learning algorithm aims at automatically learning an policy $\pi_{\theta_i}$ under the MDP framework. For single-agent MDPs with states, actions and reward defined\footnote{For the simplicity of notation, we omit subscripts of each variable to indicate the single-agent case.}, we could use standard Q-Learning or train a Deep Q Network~(DQN)~\cite{mnih2015human}, which fits a function $Q_{\phi}(q,x)$ that maps to the accumulated reward $\sum_{t=1}^{T}\gamma^t U_{t}$. To learn such an (action, state)$\rightarrow$value function, we learn the Q-table in Q-Learning or Q-network in DQN via the recursive step to minimize the following loss:
\begin{subequations}
	\label{equ:q}
	\begin{align}
	J(\phi)&=(y-Q_{\phi}(q,x))^2, \\
	y&=U_t+\gamma \max_{x_{t+1}}Q_{\phi}(q_{t+1}, x_{t+1}).
	\end{align}
\end{subequations}

Note that we need to take the expectation $\mathbb{E}_{\mathbf{q},\mathbf{x}, \mathbf{r}}$ on \eqref{equ:q} when we are using batch-based algorithms. Once we get an accurate approximation of $Q_{\phi}(q,x)$, finding the optimal policy function is essentially a network inference step:
\begin{equation}
\pi_{\theta}^*({q}_{t-K},...,{q}_t)=\arg \max_x Q_{\phi}(q,x) .
\end{equation} 

Now we discuss how to extend the single-agent setting to the multi-agent MCS system. To make our algorithm practical for real-world multiple MCS participants, we are also considering the following assumptions:
\begin{itemize}
	\item During training, each participant could observe both collective actions $\mathbf{x_t}$ and collective reward profiles $\mathbf{r_t}$;  
	\item During testing or real implementation, each participant can only use its share of information, such as observations of QoI for agents~(partially observable on $\mathbf{x_t}$);
	\item We do not assume any particular communication algorithms between participants about their sensing strategies;
	\item Each participant may face heterogeneous and stochastic QoI dynamics~(as shown in Fig.~\ref{fig:signals}).
\end{itemize}

One may directly try to extend Q-learning via \eqref{equ:q} to the multi-agent setting by finding separate $Q_{\phi_i}$ for each agent using available $q_{i}, \, x_i$. However, environment becomes non-stationary from the perspective of any individual MCS participant, since the effort level chosen by one participant would affect the payoffs of other participants. Such change can not be explained by each participant's own state-action space independently. Thus the family of Q-learning~\cite{xiao2017secure, mnih2015human} is not able to learn such non-stationary dependencies, since Q function is updating independently for each participant. Though it is possible to include all agents' decisions $\{x_{1,t},...,x_{N,t}\}$ as input for Q table/network during training process to help learn the Q function, it can not be included during testing because of the independence assumption on participant's strategies. Moreover, Q-learning is difficult to scale up to continuous actions such as the effort level in our case.

\begin{figure}[h]
	\centering
	\includegraphics[scale=0.32]{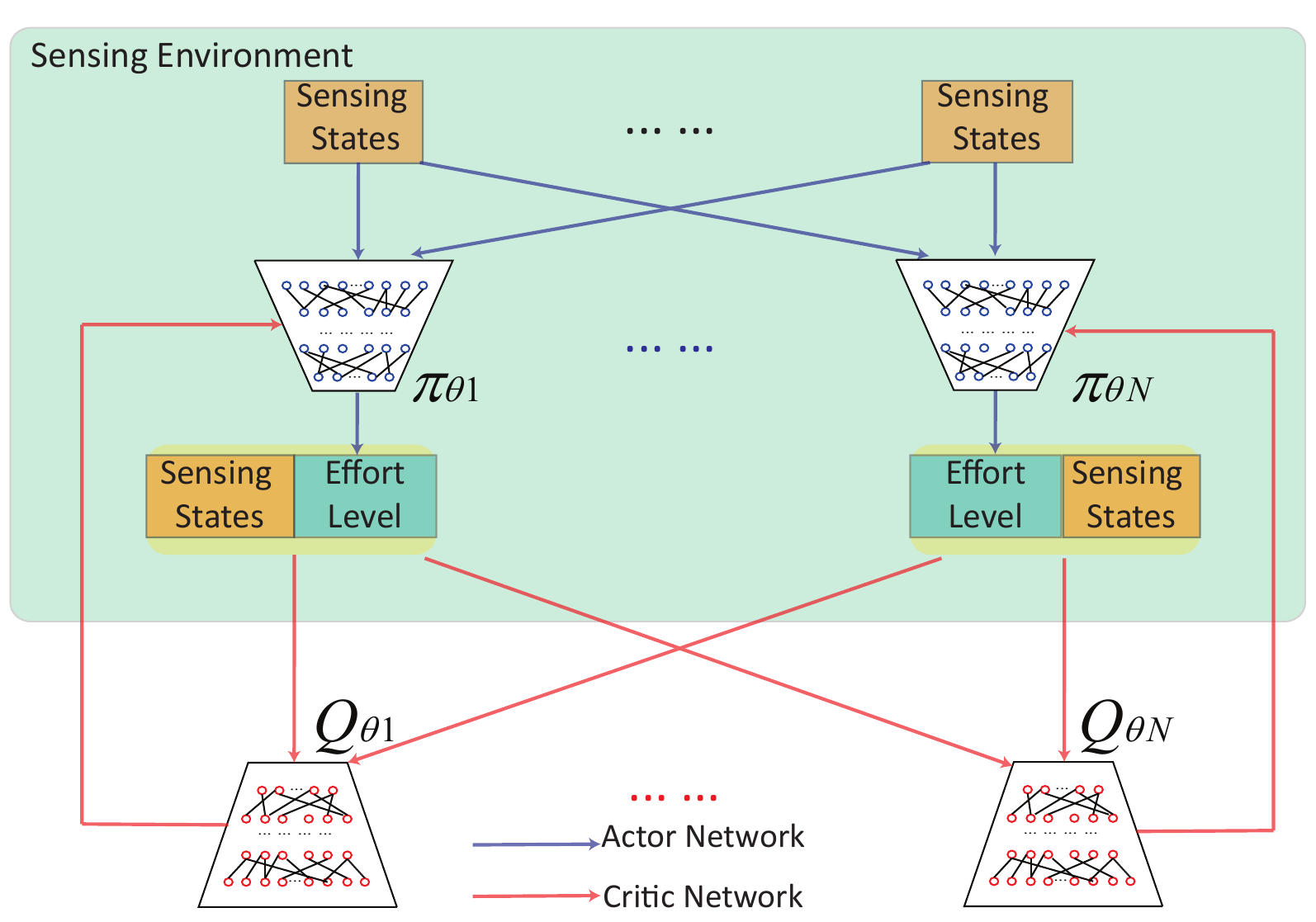}
	\caption{The schematic plot for our multi-agent reinforcement learning~(MARL) algorithm composed of a group of actor neural networks~($\pi_{\theta_i}$) and corresponding group of critic neural networks~($Q_{\phi_i}$). During training, the critic network needs to use both information from participants' QoI states and actions on sensing effort, while during implementation, the actor network only needs states information to make decisions.}
	\label{fig:rl}
\end{figure}

\subsection{IntelligentCrowd Algorithm}
To ease the learning difficulty by only using a single Q network for each agent, we adopt the actor-critic model proposed in~\cite{mnih2016asynchronous, lowe2017multi}. In the multi-agent version of actor-critic model, we are using two neural networks for each MCS participant, actor $\pi_{\theta_i}$ and critic network $Q_{\phi_i}$ to co-learn the action policy. Just as the notation suggests, the critic network is similar to the Q network in DQN, while the actor network replaces the inference step in DQN to directly get the mapping from state to the optimal effort level actions learned by it.

\begin{figure*}[t]
	\centering
	\includegraphics[scale=0.2]{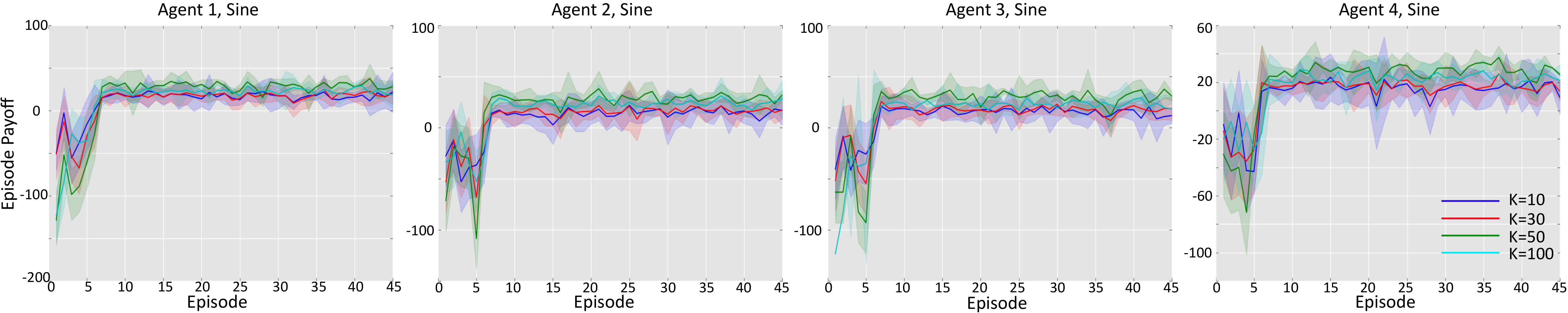}
	\caption{\small
		{
			MCS participants' rewards w.r.t training episodes with all participants under $Sine$ sensing dynamics.}
	}
	\label{fig:result_1}
\end{figure*}

In order to train both networks together, we could utilize the output from $Q_{\phi_i}$. That is, the critic could act as a judge of the policy output by the actor. By using the feedback loss from the critic network, the actor network adjusts its weights to output better decision on effort level. Once trained, the actor could directly output actions and does not need information from the critic network. Thus during training, we could add effort level from other agents to help critic learn the interactions among different agents. Mathematically, we denote the critic's input to be $(\mathbf{q},x_1,...,x_N)$, while the actor's input to be $\mathbf{q}_{t-K}^j,...,\mathbf{q}_{t}^j$ which considers $K$ step's historical VoI observations. Since critic has the full observation of each participant's effort decision during the training process, it is faced with a stationary environment, which is not subject to change of state transition function with any modification on $x_i$.

Similar to the policy gradient update approach~\cite{mnih2016asynchronous}, we can update the actor network via the gradient of policy value return:
\begin{subequations}
	\begin{align}
\nabla_{\theta_i}J(\theta_i)  =&\nabla_{\theta_i}\pi_{\theta_i}(\mathbf{q}_{t-K},...,\mathbf{q}_{t})\nabla_{x_i} Q_{\phi_i}(\mathbf{q}^j, x_1,...,x_i,...,x_N), \\
x_i=& \pi_{\theta_i}(\mathbf{q}_{t-K},...,\mathbf{q}_{t}).
\end{align}
\end{subequations}

We implement a centralized training with decentralized execution. Specifically, during training processes, each participant is able to make use of extra action information $x_j, j=1,...,N, j\neq i$ provided by other participants via the critic neural networks, while during the real implementation, they only make use of participant's own observations $\mathbf{q}_{t-K},...,\mathbf{q}_{t}$ via the actor neural networks. In this sense, there is no communication requirement for the proposed scheme.

We summarize our \emph{IntelligentCrowd} algorithm in Algorithm~\ref{algorithm1}, which is a user-centric algorithm for MCS participants to find the best sensing efforts in a multi-agent stochastic sensing environment. We also want to highlight that once \emph{IntelligentCrowd} completes the training stage in a centralized manner for the critic networks, it could be implemented in a distributed manner for each MCS participant. In the algorithm, we take the notion of episode length similarly to the MCS campaign time $T$. The actor-critic model is trained using fixed-length episode data. We also keep an experience replay buffer $\mathcal{D}$ during training, which stores the tuples $(\mathbf{q}_t, x_{i,t}, U_{i,t}, \mathbf{q}_{t+1})$ from past traces of MCS participants' effort decisions and QoI evolution. We simply use superscript $(\cdot)^j$ and $(\cdot)^{j'}$ to denote samples from $\mathcal{D}$ along with their next-step values.

 \begin{algorithm}
	\caption{\emph{IntelligentCrowd for Solving MCS Multi-agent Payoff Maximization}}
	\label{algorithm1}
	\begin{algorithmic}
		\REQUIRE Number of MCS participants $N$, discount factor $\gamma$, total time $T$, window length $K$, minibatch size $M$
		\ENSURE Neural Network weights  $\{\theta_i\},\, \{\phi_i\},\, i=1,...N$, replay buffer $\mathcal{D}$  \\
		\STATE Sample MCS initial state $\mathbf{q}_1$
		\WHILE{$\sum_{t=1}^{T}\gamma^t U_{i,t}$ has not converged}
		\FOR {$t=1,...,T$}
		\STATE \emph{\# Sample actions for replay buffer}
		\FOR {$i=1,...,N$}
		\STATE $x_{i,t}=\pi_{\theta_i}(\mathbf{q}_{t-K},...,\mathbf{q}_{t})$
		\STATE Observe payoff ${U}_{i,t}$ and next-step environment  $\mathbf{q}_{t+1}$
		\STATE $(\mathbf{q}_t, x_{i,t}, U_{i,t}, \mathbf{q}_{t+1}) \rightarrow \mathcal{D}$
		\STATE $\mathbf{q}_{t}\leftarrow \mathbf{q}_{t+1}$
		\ENDFOR
		\STATE \emph{\# Network parameters updating}
		\FOR {$i=1,...,N$}
		\STATE Sample minibatch  $(\mathbf{q}^j, x_{i}^j, U_i^j, \mathbf{q}^{j'}) $ from $\mathcal{D}$
		\STATE $y^j=U_i^j+\gamma Q_{\phi_i}(\mathbf{q}^j, x'_1,...,x'_N)|_{x'_k=\pi_{\theta_k}(\mathbf{q}_{t-K}^j,...,\mathbf{q}_{t}^j)}$
		\STATE Critic Networks update:\\ $\nabla_{\phi_i} J(\phi_i)=\frac{1}{M}\sum_{j}\nabla_{\phi_i} \left(y^j- Q_{\phi_i}(\mathbf{q}^j, x_1^j,...,x_N^j)\right)^2$
		\STATE Actor Networks update:\\
		$\nabla_{\theta_i}J(\theta_i)=\frac{1}{M}\sum_{j}\nabla_{\theta_i}\pi_{\theta_i}(\mathbf{q}_{t-K}^j,...,\mathbf{q}_{t}^j)$ \STATE $\nabla_{x_i} Q_{\phi_i}(\mathbf{q}^j, x_1^j,...x_i,...,x_N^j)|_{x_i=\pi_{\theta_i}(\mathbf{q}_{t-K}^j,...,\mathbf{q}_{t}^j)}$
		\ENDFOR
		\ENDFOR
		\ENDWHILE
	\end{algorithmic}
\end{algorithm}

%% file: results.tex
To evaluate the performance of our algorithm, we validate \emph{IntelligentCrowd} on a series of different sensing environments. We also analyze the impacts of different parameters in \emph{IntelligentCrowd}. %During the training phase, we could not only observe that each agent's payoff improves, we could also observe the relationship of rewards among different agents. 

\subsection{Simulation Setup}
\subsubsection{Neural Networks Setup}
We specifically construct four groups of actor-critic networks to learn the sensing policy upon effort level for four MCS participants. We use two-layer fully connected networks for both actor and critic networks. We vary the historical window size $K$ during our training and testing, where $K$ decides how much historical QoI data $(\mathbf{q}_t,...,\mathbf{q}_{t-K})$ are taken into consideration for both networks. Standard neural networks modeling techniques, such as batch normalization, pass-through links are adopted. We train these actor and critic networks till convergence on each agent's episode payoff and keep a fixed $T=45$ steps in all simulations. We also open-source our code on Github\footnote{github.com$/$chennnnnyize$/$Multi\_Agent-Reinforcement-Learning-for-CrowdSensing}, which is free to use for testing \emph{IntelligentCrowd}'s general performance on other MCS setups and large-scale implementations.

\begin{figure*}[!t]
	\centering
	\includegraphics[scale=0.2]{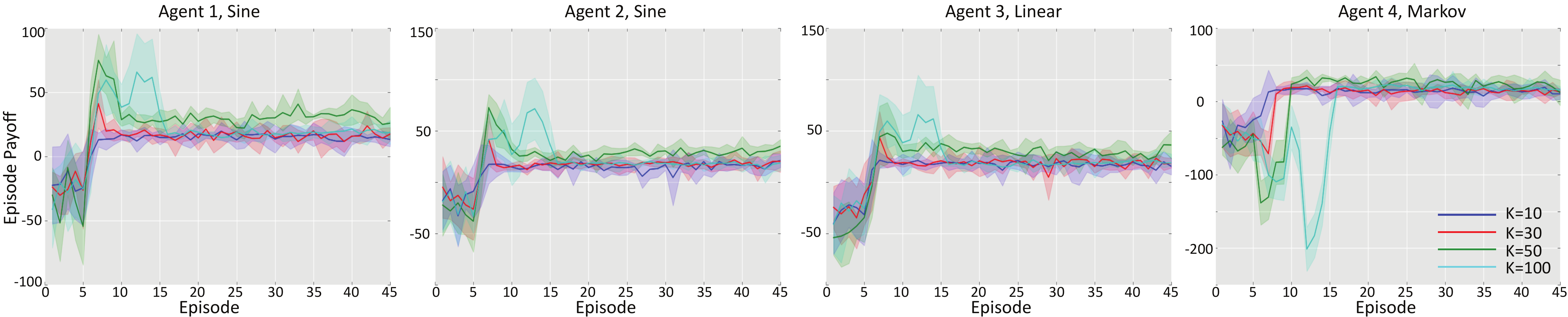}
	\caption{\small
		{
			MCS participants' payoff w.r.t training episodes under a mixture of $Sine, \; linear, \; Markov$ sensing dynamics.}
	}
	\label{fig:result_2}
\end{figure*}

\subsubsection{QoI Dynamics}
Due to the mobility of users and the
time-varying nature of sensing capabilities, we consider three
heterogeneous QoI dynamics as shown in Fig.~\ref{fig:signals}. These dynamics have been investigated in previous MCS community, or can be used to approximate more complex QoI dynamics~\cite{lei2019multiuser, he2014toward,bae2007efficiency, sutton1998reinforcement}:
\begin{itemize}
	\item \emph{Sine Dynamics:} Under this setting, the MCS participant is faced with periodic sensing signals with fixed amplitude and frequency.
	\item \emph{Linear Dynamics:} The MCS participants are receiving periodic QoI of linear strength w.r.t time.
	\item \emph{Markov chain Dynamics:} We simulate finite state space Markov chain to represent the QoI temporal evolution. 
\end{itemize}

We also make the system dynamics more challenging for \emph{IntelligentCrowd} by setting different frequencies/amplitudes/transition matrices for different MCS participants with the same dynamics. We also allow the sensed signal to be negative~(e.g., some wrong information), and each MCS participant must learn to avoid such fake information. 

\subsection{Results and Analysis}
In Fig.~\ref{fig:result_1}, we show the simulation results when $4$ agents are all under a \emph{Sine} dynamics of QoI. We plot the mean episode payoff w.r.t the training episode, along with payoff variance in $10$ runs. All $4$ participants exhibit similar learning behaviors with varying window length $K$. During the initial training episodes, all of the agents lack knowledge of the QoI dynamics and the decision patterns of other agents, so the performance is poor and participant even gets negative payoffs, which implies that MCS participants do not get sufficient reward from the service provider to compensate their costs. At around episodes 7-8, all participants start to learn a good strategy on $x_{i,t}$, and are getting greater payoffs. Such payoffs become stable as training goes on, which suggests that the neural network training for actor and critic is stable. 

We could also observe that as $K$ increases from $10$ to $50$, all of the agents are learning strategies which could get higher payoffs. This indicates that \emph{IntelligentCrowd} performs better with the aid of more information coming from past observations. However, by increasing $K$ from 50 to 100, such historical information does not help much in the decision making of sensing effort levels, but it takes more computational resources to train actor-critic networks due to the increasing dimensions of data input. 

Next, we evaluate our algorithm when MCS participants are faced with different kind of dynamics. In this setting, it is more challenging for MCS participants to make wise choices of effort levels since heterogeneous dynamics would make user interactions more complicated. As shown in Fig.~\ref{fig:result_2}, all participants are able to select effort levels to make positive payoffs when training goes to the end. Yet for participants 1, 2, and 3 who are under more ``predictable" sine or linear dynamics, they are able to make positive payoffs around 7-8 episode. Yet agent 4 under Markov decisions are starting to learn to make a positive payoff in later episodes. More interestingly, even though participants 1-3 are making high payoffs around episode 10, participant 4 starts to make good effort decisions at each timestep so as to get a portion of total payoffs from the other 3 participants and make higher payoffs in later episodes.

\begin{figure}[!t]
	\centering
	\includegraphics[scale=0.35]{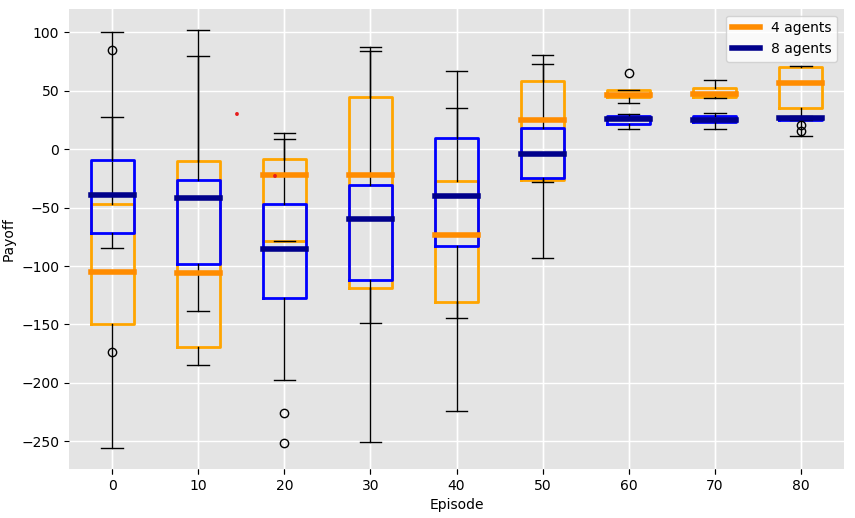}
	\caption{\small
		{
		Boxplot for MCS participants' rewards with respect to training episodes under linear dynamics for 4-agent and 8-agent task.}
	}
	\label{fig:8agent}
\end{figure}

We also implement a larger-scale MCS task using our proposed algorithm. As shown in Fig. \ref{fig:8agent}, the average reward for the 4-agent or 8-agent case both converge as more training episodes are implemented. With smaller number of agents, it takes fewer samples to learn a good RL decision maker.

In Table~I, we summarize the average accumulated rewards for four MCS participants during an episode for four types of QoI dynamics after training has finished. The mixed dynamics is the same environment we simulate in Fig.~\ref{fig:result_2}. We also observe that when we set $K=50$ in \emph{IntelligentCrowd}, MCS participants could get the highest crowdsensing payoffs.

\begin{table}
	\centering

	\label{table}
	\caption{The testing rewards on a set of different sensing dynamics with varying memory length of past sensing states. Results are compared with a MPC model using local observations.}
	\begin{tabular}{c|cccc|c| c}
	\hline
	Memory Length & 10 & 30 & 50 & 100 & MPC & Single RL \\ \hline
	Sine          &  28.25 & 25.88 & 33.52 & 28.07  & 1.87 & 24.87\\
	Linear        &  46.41  & 48.36   & 50.01   & 48.79 &  0.68 &26.75 \\
	Markov        &   35.89 & 38.88   & 44.73   & 42.31 & 1.25 & 13.77\\
	Mixed        &  22.43  &35.75    &36.91    & 27.77 & -43.52 & 11.95\\ \hline
\end{tabular}
\end{table}

We also compare our proposed method to model-predictive control (MPC) approach, which is the standard method being used for resource allocation and optimal control. We fit each agent a local model for the reward dynamics, and solve the MPC with fixed $T$ to get the predicted optimal sensing actions at each timestep. As a baseline learner, the average results for single agent RL is also considered, where each agent trains an independent actor-critic learner to learn its own sensing decisions. No communication or coordination among agents are considered in such setup.

As shown in Table I, compared to the deep RL algorithms, the MPC provides solutions with much smaller rewards in all environments. We found policy found by MPC even gets negative payoff under mixed sensing dynamics. This is possible because MPC guides each agent to achieve smaller rewards with huge sensing costs by using each agent's single observations.  On the other hand, due to the fact that we trained a multi-agent actor-critic RL decision-maker, the local learning-based agent can make use of the information provided by the critic to make decisions cooperatively. More interestingly, the MPC agent performs worst on the mixture of QoI dynamics, possibly because the linear model is unable to find a good representation of system dynamics, while the local agent are not coordinating with each other appropriately. The results shown in Table I for single-agent RL have a memory length of 100, and even though such a long period of memory is taken into account, the agent's performance is worse than MARL counterparts.

%% file: conclusion.tex
In this paper, we take MCS participants' perspective and investigate the problem of determining sensory efforts to maximize the payoff for each individual participant. We first address the challenges in modeling and decision-making since participants are faced with stochastic sensing environments and there exist complex interactions among MCS participants. Then we develop an online learning algorithm \emph{IntelligentCrowd}, which can leverage the power of deep reinforcement learning to efficiently find the best sensing decision for each participant in real time. We validate our \emph{IntelligentCrowd} algorithm by simulations in various unknown sensing environments.  Current work holds limitations on the size of problem being considered, as the proposed algorithm needs to keep the same number of actor-critic networks as the number of agents. Possible solutions may include pre-processing and clustering steps before machine learning, which holds the promise of easing the computation burden of multi-agent learning process. In the future work, we will also consider more practical setups for multi-agent learning framework, and we would like to explore the interactions between the service provider’s mechanism design and MCS participants’ decision making. Sharing representation between different agents  could be also helpful for multiple MCS participants.